\documentstyle[prl,aps]{revtex}
\draft
\begin{document}
\title{Pseudogap in the one-electron spectral functions of the attractive
Hubbard model}
\author{Massimiliano Capezzali and Hans Beck}
\address{Physics Institute \\
University of Neuch\^{a}tel \\
Rue A.L. Breguet 1 \\
2000 \underline{Neuch\^{a}tel} \\
Ch-Switzerland}
\maketitle
\widetext
\begin{abstract}
\begin{center}
\parbox{14cm}{
We calculate the one-electron Green's function of the 2D attractive Hubbard
model by coupling the electrons to pair fluctuations. The latter are 
approximated by
homogeneous amplitude fluctuations and phase correlations corresponding 
to the
XY-model. The electronic density of states shows a pseudogap at 
temperatures well
above the transition temperature $T_{C}$. For a quasi-3D system, 
a superconducting
gap emerges out of the pseudogap below $T_{C}$.
}
\end{center}
\end{abstract}
\vspace*{1.0truecm}
\noindent
The 2D attractive Hubbard model is treated by a Stratonovich-Hubbard
transformation (SHT), decoupling the interaction term by a complex pairing
field $\Delta$. The one-electron Green's function is then approximately given
by \cite{ped}:
\begin{eqnarray}
G\left(
\vec{k},z_{\nu}
\right)^{-1}=z_{\nu}-\epsilon_{\vec{k}}+\mu-\sigma\left(
\vec{k},z_{\nu}
\right)-{\left<
\Delta
\right>^{2}
\over 
z_{\nu}+\epsilon_{\vec{k}}-\mu+\sigma\left(
\vec{k},-z_{\nu}
\right)
}.
\end{eqnarray}
The expression for the self-energy
\begin{eqnarray}
\sigma\left(\vec{k},z_{\nu}
\right)=-\sum_{\vec{q}}^{}{
\sum_{z_{\alpha}}^{}{
\left<
|\Delta\left(
\vec{q},z_{\alpha}
\right)
|^{2}
\right>G\left(
\vec{k}-\vec{q},z_{\nu}-z_{\alpha}
\right)
}
},
\end{eqnarray}
involves the dynamic correlation function of the pairing field, which is
related to the one-electron propagator through the SHT.
However, rather than aiming at a self-consistent solution, we adopt a simple
form for the pairing correlations and study their influence on the
one-electron properties. Introducing amplitude and phase,
$
\Delta\left(
\vec{r},t
\right)=|\Delta\left(
\vec{r},t
\right)|e^{i\theta\left(
\vec{r},t
\right)
}
$, we make the following assumptions :\\ \\
{\it (i)} Below the temperature $T^{*}$, the amplitude fluctuations (assumed
to be space- and time-independent) are approximated by a BCS-form,
$
\left<
|\Delta|^{2}
\right>=\Phi_{0}\left(
1-{T\over T^{*}}
\right)
$, with a prefactor $\Phi_{0}$ that allows to vary their strength. The
average $\left<
\Delta
\right>
$ is zero. The strong anisotropy of underdoped compounds shows that the
coupling between the planes is weak. Thus, we model the phase fluctuations as
in a 2D-XY system. Above the critical temperature $T_{C}$, we approximate it by
a dispersionless relaxation \cite{hub,tie} :
\begin{eqnarray}
\left<
e^{i\left(
\theta\left(\vec{r},t\right)-\theta\left(\vec{0},t\right)
\right)}
\right>=e^{-\gamma t}\left(
{r_{0}\over r_{0}+r}
\right)^{\eta(T)}e^{-{r\over\xi_{+}(T)}},
\end{eqnarray}
with a Berezinskii-Kosterlitz-Thouless correlation length
$\xi_{+}(T)=
\xi_{0}e^{{b\over\sqrt{T-T_{C}}}}
$ and a relaxation frequency $\gamma$.\\ \\
{\it (ii)} Below the critical temperature $T_{C}$ we keep, for simplicity, the
same form (3) for the phase correlations, but with $\xi^{-1}=0$, corresponding
to an algebraic decay of correlations. Taking into account a non-zero coupling
between the planes (in the third dimension),
we introduce a non-zero value for the average of
$\Delta$, $\left<
\Delta
\right>^{2}=\lambda\left<
|\Delta|^{2}
\right>$, with a variable parameter $\lambda\leq 1$.\\ \\
We then evaluate $\sigma\left(\vec{k},z_{\nu}
\right)$ to lowest order by using the non-interacting
Green's function and the isotropic spectrum $\epsilon\left(
\vec{k}\right)={k^{2}\over 2m}-\mu$
in expression (2). The wave-number integration
is limited to an effective spherical Brillouin zone. The various parameters
are chosen to describe high-$T_{C}$ materials in the strongly underdoped regime,
where the pseudogap is most pronounced : the (fixed) chemical potential
$\mu={k_{F}^{2}\over 2m}$ is taken to correspond to about 0.1 charge carrier
per site and $k_{B}T^{*}=\mu$, $T_{C}={T^{*}\over 6}$.\\
On the figures, we show the electronic density of states $N(E)$ and
some spectral functions for $\Phi_{0}=0.76$ and $\gamma=0.5\mu$.
The following observations can be made :\\ \\
{\it (i)} Above $T_{C}$,
a pseudogap opens around $\mu$. Its effective width is almost
$T$-independent, in spite of the $T$-dependence of
$\left<|\Delta|^{2}
\right>$ in the self-energy. Near $T^{*}$, it is roughly V-shaped. By
approaching $T_{C}$, it becomes more U-shaped (for our model,
pairing is $s$-like) and $N\left(\mu\right)$
becomes then practically zero; the pseudogap is
delimited by two rather pronounced maxima, although $\left<\Delta\right>$
is still zero.\\ \\
{\it (ii)} Below $T_{C}$, the presence of $\left<\Delta\right>\neq 0$ produces
two new peaks: a true "superconducting gap" emerges out of the pseudogap. Its
width is given by the geometrical superposition of average and fluctuating
part of $\Delta$. The fluctuations of the latter remain visible in the form of
secondary shoulders inside the superconducting gap which approach
each other about in the same proportion as the main peaks move away
from each other.\\ \\
{\it (iii)} Our value for
$\gamma$ is relatively large.
For a smaller $\gamma$, and  
in particular in the case of critical slowing down
of phase fluctuations ($\gamma\left(T_{C}^{+}\right)=0$) \cite{hub} the
two shoulders would become secondary peaks inside the superconducting gap.\\ \\
{\it (iv)} Near $T_{C}$,
the spectral functions are doubly peaked in some wave-vector
domain around $k_{F}$ and the width of the pseudogap is given by the
separation between these two peaks, which is essentially determined by
$\left<|\Delta|^{2}
\right>$. At higher temperatures, the spectral functions have only one
peak, but their width is enhanced over essentially the same $k$-domain.
Due to this fact, the pseudogap is filling up gradually, when
$T^{*}$ is approached (from below), without changing very much
its width.\\ \\
Summarizing, we have evaluated the electronic density of states of the
attractive Hubbard model, taking into account the coupling of the
charge carriers to the fluctuating pairing field. The appearance of
a pseudogap with a $T$-independent width and the emergence of a superconducting
gap below $T_{C}$ is in agreement with experimental findings
(see for example \cite{renner}). In a future publication, we shall
present more detailed results and also compare the latter with the ones
that have been recently proposed within similar approaches \cite{others}.

FIGURE CAPTIONS
\\ \\
Figure 1 : One-electron density of states $N\left(E\right)$ above $T_{C}$, for
$T=2T_{C}$ (dot-dashed line), $T=1.5T_{C}$ (dotted line),
$T=1.2T_{C}$ (dashed line) and $T=1.01T_{C}$ (full line).\\ \\
Figure 2 : $N\left(E\right)$ below $T_{C}$ for $\lambda=0.7$ (full line) and
- for comparison - above $T_{C}$ ($T=1.01T_{C}$, dashed line).\\ \\
Figure 3 : One-electron spectral functions for $k=0.5k_{F}$ (dashed line),
$k=0.7k_{F}$ (dotted line),
$k=k_{F}$ (full line),
$k=1.2k_{F}$ (dash-dotted line),
$k=1.5k_{F}$ (heavy full line), for $T=1.01T_{C}$ and (inset)
$T=1.5T_{C}$.
\end{document}